\documentclass[twocolumn,preprintnumbers]{revtex4}%
\usepackage{amssymb}
\usepackage{amsmath}
\usepackage{graphicx}
\usepackage{epstopdf}
\usepackage{dcolumn}
\usepackage{bm}
\usepackage{xcolor}
\usepackage{amsfonts}
\usepackage{soul}
\usepackage{epstopdf}%
\providecommand{\U}[1]{\protect\rule{.1in}{.1in}}
\providecommand{\U}[1]{\protect\rule{.1in}{.1in}}
\begin{document}
\title{High temperature spectroscopy of ensembles of nitrogen vacancy centers in diamond}
\author{Mohammed Attrash}
\author{Oleg Shtempluck}
\author{Eyal Buks}
\email[corresponding author: ]{eyal@ee.technion.ac.il}
\affiliation{Andrew and Erna Viterbi Department of Electrical Engineering, Technion, Haifa
32000 Israel}
\date{\today}

\begin{abstract}
We study the spectroscopy of an ensemble of
negatively charged nitrogen-vacancy (NV$^{-}$) centers in diamond
at high temperatures between room temperature and $700 \operatorname{K}$ under high vacuum conditions. Spin resonances are studied
using optical detection of magnetic resonance (ODMR), and optical spectroscopy
is employed to study radiative transitions. Upon increasing the temperature
the intensity of radiative decay in visible and infra-red decreased. In
addition, the ODMR resonance frequencies were decreased, and the phonon line
emission shifted to higher wavelengths.  Density functional theory calculation of the zero-field splitting parameter ($D$) revealed that thermal expansion is not enough to explain the shift in the ODMR frequencies. Fitting the measured intensity of
photo-luminescence with the theoretical predictions of the Mott-Seitz model
yields the value of $0.22 \operatorname{eV}$ for the energy barrier associated with nonradiative decay.

\end{abstract}
\maketitle

\section{Introduction}

Negatively charged nitrogen vacancy (NV$^{-}$) center in diamond is considered
as a candidate for quantum technologies due to its unique physical properties,
such as long coherence time at room temperature. The technique of optical
detection of magnetic resonance (ODMR) allows monitoring spin resonances
\cite{Gruber_2012,Wolf_041001,Juraschek_127601,Pinto_1}, whereas radiative
transitions can be probed using optical spectroscopy. Many researchers
investigated the temperature dependence of luminescence of NV
\cite{Toyli_031001,Collins_2177,Chen_161903,Doherty_041201,Acosta_070801}. It
was found that the magnetic resonance frequencies were decreased after
increasing the temperature. The shift of the frequency lines is
attributed to the temperature dependence of the NV$^{-}$ zero-field splitting
parameter $D$. This dependency was explained by a combination of two
mechanisms, thermal expansion \cite{Ivady_235205} and electron-phonon
interactions \cite{Doherty_041201}. It is acclaimed that the
non-radiative processes shorten the excited-state lifetime at high
temperatures \cite{Plakhotnik_9751,Toyli_031001,Acosta_201202}. Beside shift
in frequencies, the photo-luminescence (PL) intensity was found to decrease at
high temperatures. Two models were proposed to explain PL intensity decay:
Mott-Seitz \cite{Gurney_69,Seitz_74} and Schon-Klasens
\cite{Schon_463,Wise_226,Klasens_306}. In the Mott-Seitz model, the PL
quenching at high temperature is due to the enhancement of the non-radiative
process. On the other hand, in the Schon-Klasens model, the PL decreases due
to radiationless recombination of holes and electrons at non-radiative
recombination centers. Despite the extensive studies of the NV at high
temperatures, some remaining issues were not thoroughly investigated.

\begin{figure}[t]
\par
\begin{center}
\includegraphics[width=3in,keepaspectratio]		{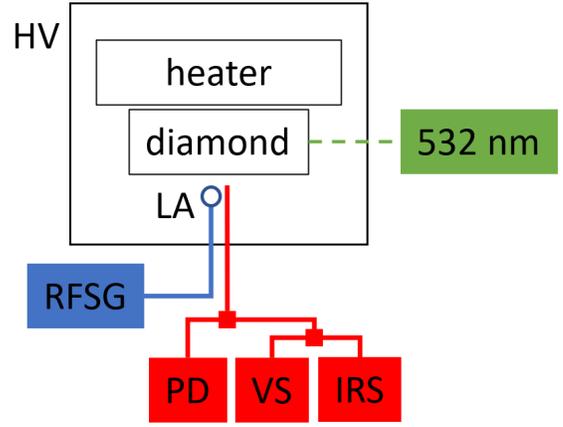}
\end{center}
\caption{Experimental setup. The diamond wafer is attached to a heater stage
inside the HV chamber. Both a MW LA and a multi-mode optical fiber are
attached to a copper-made finger cooled by liquid nitrogen. Vacuum feedthroughs are employed to connect both MW LA and optical fiber to room temperature instruments. The bare end of the optical fiber is positioned in front of the diamond wafer. The fiber
is split to 3 components (using fiber couplers) , which are connected to the PD, VS and IRS. \color{black} The
NV$^{-}$ is excited by a $532\operatorname{nm}$ laser delivered via free
space. The optical propagation direction is parallel to the [110] plane. The magnetic field is applied by a cylindrical neodymium magnet.}%
\label{FigSetupS}%
\end{figure}

\begin{figure*}[ptb]
\begin{center}
\includegraphics[width=6.2in,keepaspectratio]		{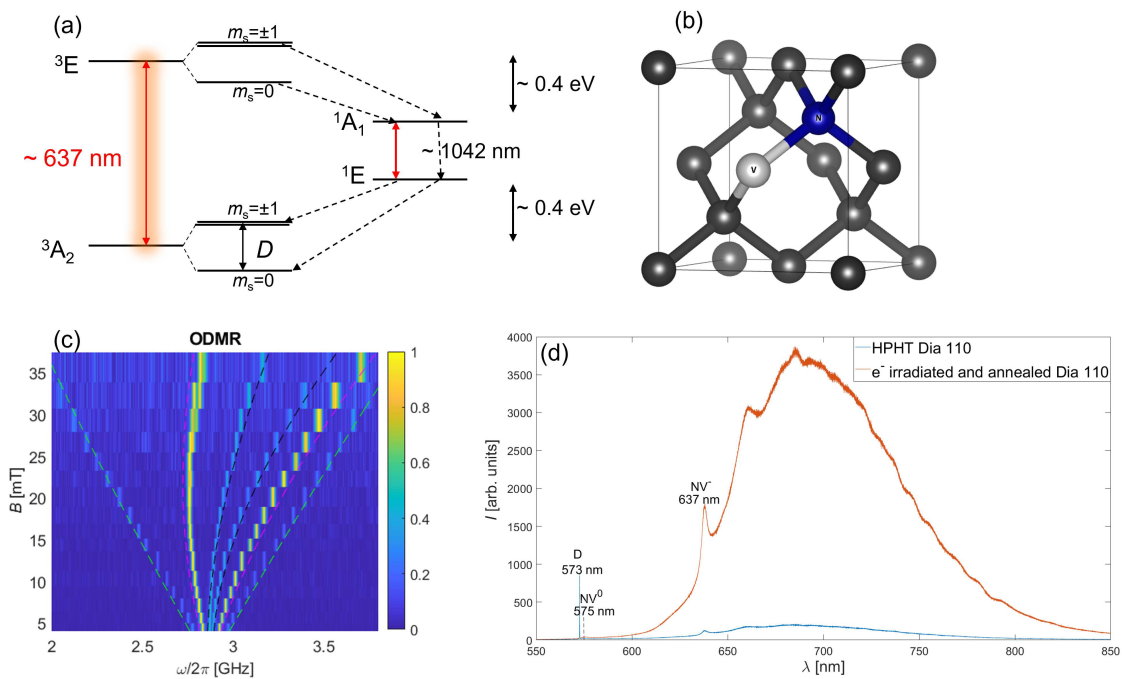}
\end{center}
\caption{(a) Schematics of the NV$^{-}$ electronic structure, the $^{3}
\mathrm{A}_{2}$ and $^{3}\mathrm{E}$ are the triplet ground and excited
states, respectively, while $^{1}$A$_{1}$ and $^{1}$E are singlet states,
which involved in the intersystem crossing. The solid lines represent the
optical transitions, while the dashed lines represent the non-radiative
transitions. (b) NV unit cell structure, where carbon atoms are in gray,
nitrogen in blue, and vacancy in white. (c) ODMR plot at room temperature. The color bar indicates the normalized ODMR intensity. The dashed lines represent the resonances
related to spin transitions from $m_{\mathrm{s}}=\pm1$ to $m_{\mathrm{s} }=0$
for the four different NV$^{-}$ center directions (the resonances corresponding to two
directions are nearly degenerate). (d) PL measurement of
HPHT diamond [110] before and after electron
irradiation and annealing. For HPHT diamond, two features are observed: at 573
nm, which is related to diamond Raman peak
\cite{Mildren_18950,Chandran_221602}, and at 637 nm, which is related to
NV$^{-}$ zero-phonon line. After electron irradiation and annealing, a very
low peak intensity at 575 nm can be observed which is related to NV$^{0}$
centers in diamond. In addition, the phonon line intensity (600 - 850 nm)
increased.}%
\label{FigODMR}%
\end{figure*}

In this work we employ both ODMR in the microwave (MW) band, and optical
spectroscopy in the optical band of $500-1500 \operatorname{nm}$, to study ensembles of NV$^{-}$ centers in a high vacuum (HV)
system at different temperatures ($300 - 550 \operatorname{K}$). In addition, we investigate the PL as a function of temperature and calculate the energy barrier according to Mott-Seitz model of ensembles of NV$^-$ centers. \color{black} The comparison between our experimental results and theory yields partial agreement.

\section{Experimental setup}

Type Ib high pressure high temperature (HPHT) single crystal diamond with a
nitrogen concentration lower than $200$ ppm was laser-cut along the
[110] plane, polished and irradiated with $2.8
\operatorname{MeV}$ electrons at a doze of $8\times10^{18} \operatorname{cm}
^{-2}$, and annealed at $900^{\circ}$C for 2 hours. The estimated NV$^{-}$
concentration is $3.3\times10^{17} \operatorname{cm}^{-3}$
\cite{Alfasi_214111}. The diamond was cleaned in an acid mixture of
Perchloric, Sulfuric, and Fuming Nitric acid for 1 hour.

The experimental setup is schematically shown in Fig. \ref{FigSetupS}. The
vacuum system, which has a base pressure of $1\times10^{-6}
\operatorname{Torr}$, includes an ultra HV heater (HeatWave
Labs Inc., model 101491) connected to a thermocouple (type K) and a
temperature controller (model 101303), MW loop antenna (LA), and a bare multi-mode
optical fiber pointing to the diamond \color{black} to collect PL emitted from the diamond. Liquid nitrogen was
flowed in a metal finger attached to the LA and the optical fiber to avoid
annealing damage. The optical fiber is split (1x2 Step-Index Multimode Fiber Optic Couplers - Thorlabs) \color{black} to 3 components, which are
connected to: photo-diode (PD) (Thorlabs, PDA100A), visible spectrometer (VS)
(Thorlabs, CCS175, $500-1000 \operatorname{nm}$), and infrared spectrometer
(IRS) (Ibsen photonics, ROCK NIR $900-1700 \operatorname{nm}$). The relative
optical power delivered to the PD, VS and IRS is $0.5$, $0.25$ and $0.25$,
respectively. The spectrometers allow monitoring both the triplet-triplet transition at wavelength of $637
\operatorname{nm}$ and the singlet-singlet transition at $1042 \operatorname{nm}$ [see the schematic energy level diagram
shown in Fig. \ref{FigODMR}(a)].

The NV$^{-}$ centers were excited by a $532 \operatorname{nm}$ green laser
delivered via free space. The LA is connected to a radio frequency signal
source (RFSG), and the MW amplitude was modulated by a $151 \operatorname{Hz}$
sine wave. The PD is attached to a $600 \operatorname{nm}$ long-pass filter,
and its signal was demodulated by a lock-in amplifier. To apply magnetic field
on the diamond, a cylindrical neodymium magnet is positioned outside the
vacuum chamber using a motorized stage. To reduce laser intensity
fluctuations, we used proportional integral derivative (PID) \color{black} controller (SIM 960 -
Stanford Research Systems). A beam splitter in front of the laser together with a PD are used to generate the PID input signal.  The longtime optical intensity stability is improved by more that a factor of 10 (compared with the stability obtained without the PID). \color{black}

\section{ODMR}

Ignoring the hyper-fine interaction, the NV$^{-}$ spin triplet ground state
Hamiltonian $\mathcal{H}_{\mathrm{NV}}$ is given by%
\begin{equation}
\frac{\mathcal{H}_{\mathrm{NV}}}{\hbar}=\frac{D\mathcal{S}_{z}^{2}}{\hbar^{2}%
}-\frac{\gamma_{\mathrm{e}}\Vec{B}\cdot\mathcal{S}}{\hbar}+\frac
{E_{\mathrm{NV}}\left(  S_{+}^{2}+S_{-}^{2}\right)  }{2\hbar^{2}}\;,
\label{H_NV}%
\end{equation}
where $D$ and $E_{\mathrm{NV}}$ are the zero field
splitting parameters, which equale to $2\pi
\times2.88\operatorname{GHz}$ and $2\pi\times10\operatorname{MHz}$
at $3.6\operatorname{K}$, respectively \cite{Alfasi_214111},
$\mathcal{S}$ is the total spin operator $\mathcal{S}=\mathcal{S}_{x}\hat
{x}+\mathcal{S}_{y}\hat{y}+\mathcal{S}_{z}\hat{z}$ , $\gamma_{\mathrm{e}}$
$=2\pi\times28.03\operatorname{GHz}\operatorname{T}^{-1}$ is the electron spin
gyromagnetic ratio, $\Vec{B}$ is the magnetic field, and $\mathcal{S}_{\pm
}=\mathcal{S}_{x}\pm i\mathcal{S}_{y}$. In this Hamiltonian we neglect also the on-axis electric field and some strain components which contribute to the $\mathcal{S}_z^2$ term and the off-axis electric field and other strain components which contribute to $\mathcal{S}_x$ and $\mathcal{S}_y$ terms \cite{Udvarhelyi_075201}. \color{black} Under continuous
laser excitation, the NV$^{-}$ is polarized to the spin state
$m_{\mathrm{s}}=0$, which has a brighter PL \cite{Doherty_1}.

Figure \ref{FigODMR}(a) shows a schematic energy level diagram of an NV$^{-}$
center. The structure of nitrogen vacancy center in diamond is shown in Fig.
\ref{FigODMR}(b). Measured ODMR of our sample as a function of magnetic field
is shown in Fig. \ref{FigODMR}(c). The eight lines that are
observed in this figure represent the resonances related to spin transitions
from $m_{\mathrm{s}}=\pm1$ to $m_{\mathrm{s}}=0$ for the four different
NV$^{-}$ center directions. The dashed lines represent the theoretically
calculated frequencies of the same resonances, which are obtained by
numerically diagonlizing the Hamiltonian $\mathcal{H}_{\mathrm{NV}}$
(\ref{H_NV}). Note that two pairs of resonances, which are represented by the
pink dash lines, are nearly degenerate (note that their ODMR signal is nearly
double the signal of the other NV directions). Figure
\ref{FigODMR}(d) reveals the PL measurement of the HPHT diamond sample before
and after electron irradiation and annealing. The measurement was recorded by
Micro-Raman spectrometer (LABRAM HR, HORIBA, Jobin Yvon) using
$532\operatorname{nm}$ wavelength laser at room temperature (laser
power was lower than 10 mW). The HPHT diamond reveals a peak at
$573\operatorname{nm}$ related to diamond Raman peak
\cite{Mildren_18950,Chandran_221602}, and another low intensity peak at
$637\operatorname{nm}$ that is related to the zero phonon line of NV$^{-}$
[see Fig. \ref{FigODMR}(a)] \cite{Prawer_2537,Rondin_115449}. After electron
irradiation and annealing process, the intensity of NV$^{-}$ zero phonon line
increased, beside the PL intensity increase in the entire range of
$650-850\operatorname{nm}$. The intensity of the NV$^{0}$ zero phonon line is
very low in the treated diamond sample. The reason for relatively low
NV$^{0}$ peak intensity may be related to low laser power, as was found in Ref. \cite{Manson_1705}.

\section{Zero field splitting $D$}

\begin{figure}[ptb]
\begin{center}
\includegraphics[width=3.6in]{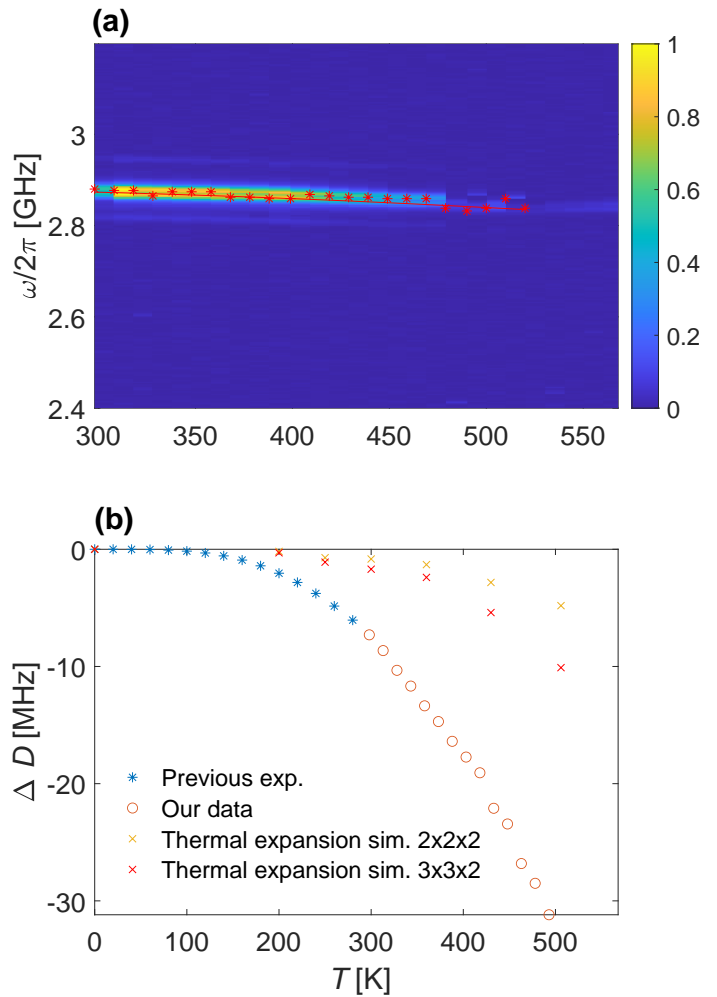}
\end{center}
\caption{(a) and (b) ODMR as a function of temperature recorded at
magnetic field of $0$ and $10\operatorname{mT}$, respectively. The ODMR
frequencies decrease as temperature increases due to the decrease of $D$.
Fitting of $D$ was done according to the parameters presented in Ref.
\cite{Toyli_031001}. (c) The shift $\Delta D$ in $D$ as a function of
temperature. Note that $D$ is assumed to be independent on the
externally applied magnetic field for all fitting lines in (a) and (b). The
previous experimental data [see blue stars in (c)] are from
\cite{Feshchenko_1609_06519,Doherty_041201}.}%
\label{FigPLvsT}%
\end{figure}

To numerically investigate the shift in the zero field splitting $D$, density
functional theory (DFT) \cite{Hohenberg_B864,Kohn_A1133} was used as
implemented in the QUANTUM ESPRESSO software \cite{Giannozzi_395502}. The
defect structure was modeled with $3\times3\times2$ and $2\times2\times2$ super-cell of diamond (144 and 64 atoms, respectively) \color{black} and $\Gamma$
point sampling of Brillouin zone. The calculation were performed with
projector augmented-wave (PAW) method \cite{Blochl_17953} and
Perdew-Burke-Ernzerhof (PBE) exchange-correlation functional
\cite{Perdew_3865} with a plane-wave cutoff of $300$ Ry. To investigate the
thermal expansion effect on $D$, the lattice length of the structure was
chosen to suit the temperature as found in previous experiments
\cite{Skinner_39,Sato_092102}, and the structure was relaxed at constant
lattice length until the total force on the atoms was lower than
$26\times10^{-3} \operatorname{eV} \operatorname{\text{\AA}}^{-1}$. The zero
field splitting $D$ is believed to be related to spin-spin interactions of
unpaired electrons \cite{Lenef_13441}. Its tensor components $D_{ab}$ are
given by \cite{harriman2013theoretical}%
\begin{align}
D_{ab}  &  =\frac{\left(  \gamma_{\mathrm{e}}\hbar\right)  ^{2}}%
{2S(2S-1)}\frac{\mu_{0}}{4\pi}\nonumber\\
&  \times\sum_{i<j}\chi_{ij}\left\langle \psi_{ij}\right\vert \frac
{r^{2}\delta_{ab}-3r_{a}r_{b}}{r^{5}}\left\vert \psi_{ij}\right\rangle
\;,\nonumber\\
&
\end{align}
where $a,b\in\left\{  x,y,z\right\}  $ are Cartesian indices, $\mu_{0}$ is the
magnetic permeability of free space, the summation includes all pairs of
occupied Kohn-Sham orbitals, $\chi_{ij}=\pm1$ for parallel and anti-parallel
spins, respectively, $S$ is the spin number, and $\psi_{ij}(r,r^{\prime})$ is
$2\times2$ determinant which can be written as $\psi_{ij}(r,r^{\prime
})=2^{-1/2}[\psi_{i}(r)\psi_{j}(r^{\prime})-\psi_{i}(r^{\prime})\psi_{j}(r)]$.
A Python package named PyZFS was used to calculate $D_{ab}$ \cite{Ma_2160}.

According to our DFT calculations, the $D$ at zero temperature equals to $2.84
\operatorname{GHz}$ for $3\times3\times2$ super-cell ($2.77
\operatorname{GHz}$ for $2\times2\times2$ super-cell) while the experimental value at $3.6 \operatorname{K}$ equals to $2.88 \operatorname{GHz}$. For
comparison to other DFT results, $D$ was found in the range of $\left[
2.7,3\right]  \operatorname{GHz}$
\cite{Bhandari_014115,Whiteley_490,Ghosh_043801}. The shift of $D$ at a given
temperature, relative to its value at zero temperature, is denoted by $\Delta
D$. Some studies calculated the shift $\Delta D$ according to the thermal
expansion at temperatures below $300 \operatorname{K}$
\cite{Doherty_041201,Ivady_235205}. To account for our experimental results,
we calculate $\Delta D$ based on the thermal expansion in the range from
absolute zero to $500 \operatorname{K}$.

\begin{figure}[ptb]
\begin{center}
\includegraphics[height=5.0in, width=3.0in]		{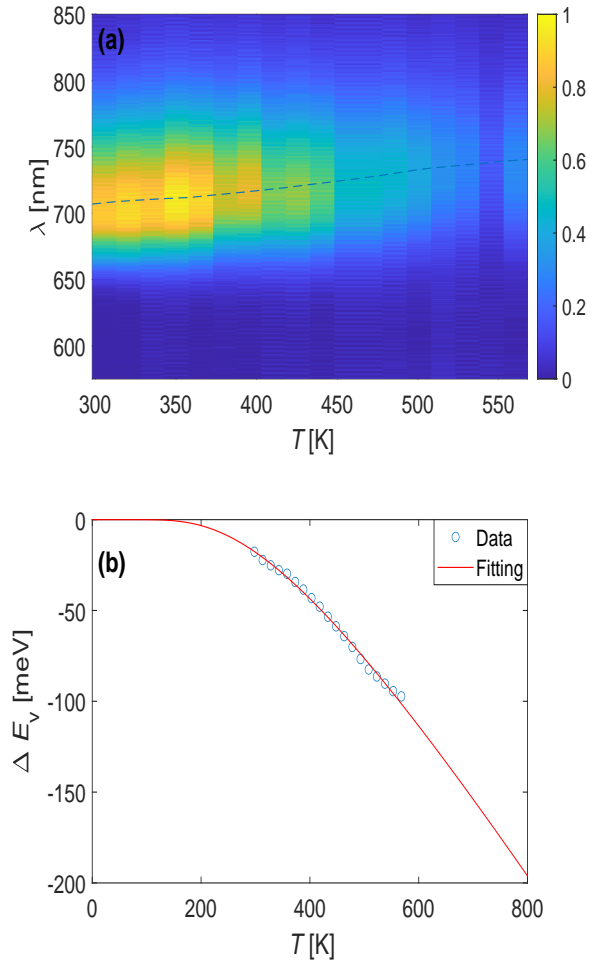}
\end{center}
\caption{(a) Optical emission spectra of NV$^{-}$ at different temperatures.
The peak intensity shifted to higher wavelength at increased temperatures. (b)
Peak shift $\Delta E_{\mathrm{V}}$ as a function of the temperature $T$.}%
\label{FigNIR}%
\end{figure}

In order to investigate the temperature effect on the ODMR, the diamond was
measured \color{black} at different temperatures under constant magnetic field. The ODMR at
temperatures below $520 \operatorname{K}$ [see Fig \ref{FigPLvsT}(a)] reveals
resonance lines related to NV$^{-}$. Upon increasing the temperature, the
resonance frequencies and their intensity were decreased, and above $520
\operatorname{K}$ the resonance lines could not be resolved. It is expected
that the shift in resonance lines is due to the decrease of $D$. The measured
shift $\Delta D$ in $D$ [see Fig. \ref{FigPLvsT}(b)] at $500 \operatorname{K}$
equals to $-32 \operatorname{MHz}$ and the slope $\mathrm{d}D/\mathrm{d}T$ is
of order of $-100 \operatorname{kHz} \operatorname{K}^{-1}$, in agreement with
the value reported in \cite{Toyli_031001}. The normalized slope $\mathrm{d}D/(D\mathrm{d}T)$ \color{black} is $-4.3\times10^{-5} \operatorname{K}^{-1}$, which is higher
than what was found in experiments carried out at lower temperatures, for
which the value of $-2.61\times10^{-5} \operatorname{K}^{-1}$ has been
measured \cite{Acosta_070801,Chen_161903}.

The experimental results of $D$ at temperatures lower than $300
\operatorname{K}$ were taken from other reports
\cite{Chen_161903,Doherty_041201}. As can be observed from simulation results,
$D$ decreases as the temperature increases. The shift in $D$ at $500
\operatorname{K}$ compare to its value at $300 \operatorname{K}$ according to
our calculations is $-10 \operatorname{MHz}$ ($-4 \operatorname{MHz}$ for the $2\times2\times2$ super-cell). As can be seen from the plot in
Fig. \ref{FigPLvsT}(b), the thermal expansion is not enough to explain the
shift in $D$. The unit cell length $a$ of diamond at $298 \operatorname{K}$
and $506 \operatorname{K}$ equals to $3.5668 \operatorname{\text{\AA}}$ and
$3.5680 \operatorname{\text{\AA}}$, respectively \cite{Skinner_39}, therefore, \color{red}
$\mathrm{d} a/(a\mathrm{d}T) \simeq1.6\times10^{-6} \operatorname{K}^{-1}$ \color{black} in
this range. This value is too small to account for the above-mentioned
experimental finding that $\mathrm{d}D/(D\mathrm{d}T)$ \color{black} is~ $-4.3\times10^{-5}
\operatorname{K}^{-1}$ in the same range.

According to a first-principles calculation reported in \cite{Tang_2205_02791}%
, the shift in $D$ at $500 \operatorname{K}$ is $-18 \operatorname{MHz}$,
while the shift in our data is $-32 \operatorname{MHz}$. Good agreement is
found between our experimental results, and the ones reported in Ref.
\cite{Toyli_031001}, in which the zero-field splitting parameter $D$ is
expressed as a power series of the temperature $T$.

\begin{figure}[ptb]
\centering
\par
\begin{center}
\includegraphics[width=3.0 in]		{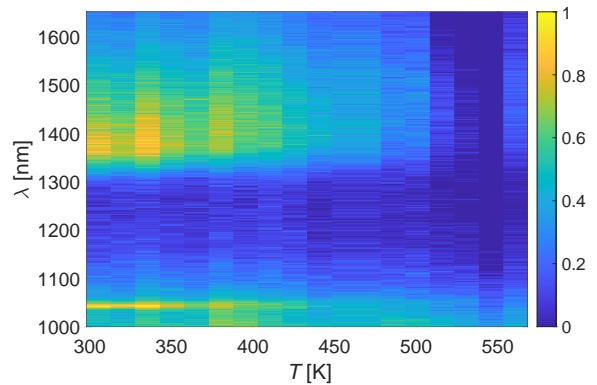}
\end{center}
\caption{IR emission as a function of temperature $T$ and wavelength $\lambda
$. The peak at 1042 nm, which represent the radiative decay in the
intermediate singlet states $^{1}$A$_{1}\xrightarrow{}^{1}$E, can be observed
at temperatures below $450 \operatorname{K}$.}%
\label{FigIR}%
\end{figure}

\begin{figure}[ptb]
\begin{center}
\includegraphics[width=3.0 in]		{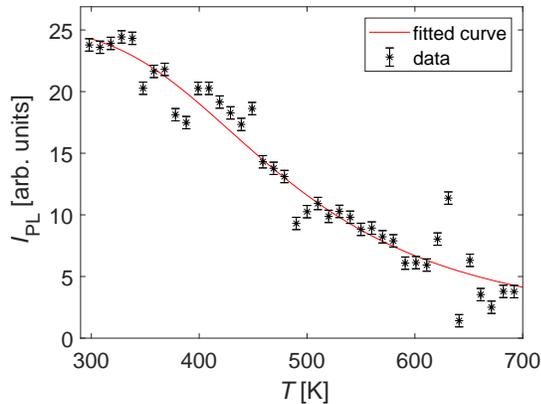}
\end{center}
\caption{PL intensity $I_{\mathrm{PL}}$ as a function of
temperature $T$. The red solid line is calculated using Eq. (\ref{I_PL}).}%
\label{FigPL}%
\end{figure}

The ODMR signal intensity can be derived from the rate equation
that governs the time evolution of the spin polarization (see appendix A of
Ref. \cite{Alfasi_214111}). As can be seen from Fig. \ref{FigPLvsT}, ODMR
signals could not be experimentally resolved above a temperature of about $520
\operatorname{K}$. On the other hand, PL signals were measured at higher
temperatures up to about $700 \operatorname{K}$ (see Fig. \ref{FigPL}). This
observation suggests that the dominant mechanism responsible for the observed
ODMR signal drop with temperature [see Fig. \ref{FigPLvsT}(a)], is dephasing, which enhances spin transverse relaxation, and consequently disables ODMR above $520 \operatorname{K}$.

\section{Optical spectroscopy}

The fluorescence emission spectra of NV$^{-}$ (measured by the VS) at
different temperatures is shown in Fig. \ref{FigNIR}. The fluorescence can be
resolved at temperatures up to $520\operatorname{K}$. The spectra is shifted
to longer wavelengths at higher temperatures. The shift in wavelength at which
the emission intensity is maximized, which is denoted by $\lambda
_{\mathrm{max}}$, equals to $hc/60\operatorname{meV}$ at $500\operatorname{K}%
$, where $h$ is the Planck constant and $c$ is the speed of light, and thus
$\mathrm{d}\lambda_{\mathrm{max}}/(\lambda_{\mathrm{max}}\mathrm{d}T)\simeq3.1\times
10^{-4}\operatorname{K}^{-1}$. Fitting of the phonon shift was performed according to \cite{Liu_3391}
\begin{equation}
    \Delta E_v(T)= \frac{-2A}{\exp \left( {\frac{\hbar \omega _0 }{2k_{\mathrm{B}}T}} \right)-1}\;,
\end{equation}
where $A$ depends on the diamond dispersion curves details, $k_{\mathrm{B}}$ is the Boltzmann’s constant, and $\omega_{0}$ is the zone-center phonon frequency. The fitting yields that $\hbar\omega_{0}=173 \pm 10 \operatorname{meV}$ and $A=246 \pm 34 \operatorname{meV}$. Note that the zone-center optical phonon Raman energy of diamond equals to $165 \operatorname{meV}$ ($1332 \operatorname{cm^{-1}}$) \cite{Solin_1687}.

\color{black} The IR spectra of the NV$^{-}$ ensemble (measured by the IRS) at different
temperatures is shown in Fig. \ref{FigIR}. The peak at $1042 \operatorname{nm}%
$, which is associated with radiation decay from the $^{1}\mathrm{A}_{1}$ to
$\mathrm{E}_{1}$ levels [see Fig.\ref{FigODMR}(a)], is observed at room
temperatures. Upon heating the diamond, the $1042 \operatorname{nm}$ peak
intensity decreased and became unresolved at temperatures above $450
\operatorname{K}$. Temperature dependence of the $^{1}\mathrm{A}_{1}$ to
$\mathrm{E}_{1}$ transition wavelength has been observed in
\cite{Doherty_041201}. However, this dependency cannot be resolved in our
measurements due to limited IRS resolution. Another feature can be observed at
$1350 \operatorname{nm}$. We find that this feature disappears when a  808 nm long pass filter is added, i.e. it originates from second order diffraction (note that $1350/2=675$). \color{black} The singlet NV$^{-}$ transition (from $^{1}\mathrm{A}_{1}$ to $\mathrm{E}_{1}$
states) wavelength is found to be less sensitive to temperature compare to the
triplet transition (from $^{3}\mathrm{A}_{2}$ to $^{3}\mathrm{E}$ states) wavelength.

Finally, we discuss the PL intensity and energy barrier for the nonradiative
process, which is denoted by $U_{\mathrm{b}}$. The PL intensity
$I_{\mathrm{PL}}$ as a function of the temperature $T$ is shown in Fig.
\ref{FigPL}. In the Mott-Seitz model this dependency of
$I_{\mathrm{PL}}$ on $T$ is given by [see Eq. (2) of Ref. \cite{Akselrod_3364}%
, and Eq. (5) of Ref. \cite{Chithambo_1880}]%
\begin{equation}
I_{\mathrm{PL}}=\frac{I_{0}}{1+C\exp\left(  -\frac{U_{\mathrm{b}}%
}{k_{\mathrm{B}}T}\right)  }\;, \label{I_PL}%
\end{equation}
where $I_{0}$ is the zero temperature intensity, and $C$ is a constant. Fitting the measured PL
intensity $I_{\mathrm{PL}}$ with Eq. (\ref{I_PL}) yields $U_{\mathrm{b}}$ $=0.22\pm0.05\operatorname{eV}$ and $C=200$.
Note that $U_{\mathrm{b}}$ was found in ref. \cite{Toyli_031001} to be
$0.48\operatorname{eV}$. The deviation between this result \cite{Toyli_031001}, which was obtained with a single NV$^{-}$ center, and our extracted value for the energetic barrier, can perhaps be attributed to impurities near the single NV center that was studied in Ref. \cite{Toyli_031001}. The effect of nearby impurities on coherence lifetime of NV$^{-}$ centers has been studied in \cite{Maze_644}.

\section{Summary}

In this study we investigated the optical properties of an ensemble of
NV$^{-}$ centers in diamond as a function of temperature. The NV$^{-}$
emission in the visible ($\sim700\operatorname{nm}$) and IR ($\sim
1042\operatorname{nm}$) regions decreased upon increasing the temperature, and
the phonon line wavelength at maximum intensity of the NV$^{-}$
($650-800\operatorname{nm}$) increased. The $D$ parameter decreased upon
increasing the temperature, and the thermal expansion model is not enough to
explain this behaviour. By applying the Mott-Seitz model we find that the
energy barrier for nonradiative decay $U_{\mathrm{b}}$ is in the range $\left[  0.17,0.27\right] \operatorname{eV}$.

This work was supported by the Israeli science foundation, the Israeli
ministry of science, and by the Technion security research foundation. We
would like to thank Joan Adler for
providing computational facilities. 

\bibliographystyle{ieeepes}
\bibliography{Eyal_Bib}

\end{document}